\documentclass[12pt,preprint]{aastex}
\usepackage{graphicx}

\shorttitle{HiRes electron-neutrino limit}
\shortauthors{Abbasi et al.}

\begin{document}
\title{An upper limit on the electron-neutrino flux \\
from the HiRes detector}

\author{
R.~U.~Abbasi,\altaffilmark{1}
T.~Abu-Zayyad,\altaffilmark{1}
M.~Allen,\altaffilmark{1}
J.~F.~Amann,\altaffilmark{2}
G.~Archbold,\altaffilmark{1}
K.~Belov,\altaffilmark{1}
J.~W.~Belz,\altaffilmark{3}
S.~Y.~Ben~Zvi,\altaffilmark{4}
D.~R.~Bergman,\altaffilmark{5}
A.~Biesiadecka,\altaffilmark{5}
S.~A.~Blake,\altaffilmark{1}
J.~H.~Boyer,\altaffilmark{4}
O.~A.~Brusova,\altaffilmark{1}
G.~W.~Burt,\altaffilmark{1}
C.~Cannon,\altaffilmark{1}
Z.~Cao,\altaffilmark{1}
W.~Deng,\altaffilmark{1}
Y.~Fedorova,\altaffilmark{1}
J.~Findlay,\altaffilmark{1}
C.~B.~Finley,\altaffilmark{4}
R.~C.~Gray,\altaffilmark{1}
W.~F.~Hanlon,\altaffilmark{1}
C.~M.~Hoffman,\altaffilmark{2}
M.~H.~Holzscheiter,\altaffilmark{2}
G.~Hughes,\altaffilmark{5}
P.~H\"{u}ntemeyer,\altaffilmark{1}
D.~Ivanov,\altaffilmark{5}
B.~F~Jones,\altaffilmark{1}
C.~C.~H.~Jui,\altaffilmark{1}
K.~Kim,\altaffilmark{1}
M.~A.~Kirn,\altaffilmark{3}
B.~C.~Knapp,\altaffilmark{4}
E.~C.~Loh,\altaffilmark{1}
M.~M.~Maestas,\altaffilmark{1}
N.~Manago,\altaffilmark{6}
E.~J.~Mannel,\altaffilmark{4}
L.~J.~Marek,\altaffilmark{2}
K.~Martens,\altaffilmark{1}
J.~A.~J.~Matthews,\altaffilmark{7}
J.~N.~Matthews,\altaffilmark{1}
S.~A.~Moore,\altaffilmark{1}
A.~O'Neill,\altaffilmark{4}
C.~A.~Painter,\altaffilmark{2}
L.~Perera,\altaffilmark{5}
K.~Reil,\altaffilmark{1}
R.~Riehle,\altaffilmark{1}
M.~D.~Roberts,\altaffilmark{7}
D.~Rodriguez,\altaffilmark{1}
M.~Sasaki,\altaffilmark{6}
S.~R.~Schnetzer,\altaffilmark{5}
L.~M.~Scott,\altaffilmark{5}
M.~Seman,\altaffilmark{4}
G.~Sinnis,\altaffilmark{2}
J.~D.~Smith,\altaffilmark{1}
R.~Snow,\altaffilmark{1}
P.~Sokolsky,\altaffilmark{1}
C.~Song,\altaffilmark{4}
R.~W.~Springer,\altaffilmark{1}
B.~T.~Stokes,\altaffilmark{1}
S.~R.~Stratton,\altaffilmark{5}
J.~R.~Thomas,\altaffilmark{1}
S.~B.~Thomas,\altaffilmark{1}
G.~B.~Thomson,\altaffilmark{5}
D.~Tupa,\altaffilmark{2}
L.~R.~Wiencke,\altaffilmark{1}
A.~Zech\altaffilmark{5}
and
X.~Zhang\altaffilmark{4}
}


\altaffiltext{1}{University of Utah,
Department of Physics and High Energy Astrophysics Institute,
Salt Lake City, UT 84112, USA.}
\altaffiltext{2}{Los Alamos National Laboratory,
Los Alamos, NM 87545, USA.}
\altaffiltext{3}{University of Montana, Department of Physics and Astronomy,
Missoula, MT 59812, USA.}
\altaffiltext{4}{Columbia University, Department of Physics and
Nevis Laboratories, New York, NY 10027, USA.}
\altaffiltext{5}{Rutgers --- The State University of New Jersey,
Department of Physics and Astronomy, Piscataway, NJ 08854, USA.}
\altaffiltext{6}{University of Tokyo,
Institute for Cosmic Ray Research,
Kashiwa City, Chiba 277-8582, Japan.}
\altaffiltext{7}{University of New Mexico,
Department of Physics and Astronomy,
Albuquerque, NM 87131, USA.}

\begin{abstract}
Air-fluorescence detectors such as the High Resolution Fly's Eye (HiRes)
detector are very sensitive to upward-going, Earth-skimming ultrahigh
energy electron-neutrino-induced showers.
This is due to the relatively large interaction cross sections of these
high-energy neutrinos and to the Landau-Pomeranchuk-Migdal (LPM) effect.
The LPM effect causes a significant decrease in the cross sections for
bremsstrahlung and pair production, allowing charged-current
electron-neutrino-induced showers occurring deep in the Earth's crust
to be detectable as they exit the Earth into the atmosphere.
A search for upward-going neutrino-induced showers in the HiRes-II
monocular dataset has yielded a null result.
From an LPM calculation of the energy spectrum of charged particles
as a function of primary energy and depth for electron-induced showers
in rock, we calculate the shape of the resulting profile of these 
showers in air.
We describe a full detector Monte Carlo simulation to determine the
detector response to upward-going electron-neutrino-induced cascades and
present an upper limit on the flux of electron-neutrinos.
\end{abstract}


\section{Introduction\label{sec:intro}}
We report on a search for upward-going electron-neutrino showers in 
the High-Resolution Fly's Eye II data set, and on the upper limit on the
flux of $\nu _e$ set by the HiRes-II detector.
The HiRes project has been discussed previously 
\citep{Abu-Zayyad_ICRC_1999a,Boyer_NIMa_2002}; 
the detector is an air-fluorescence detector located on 
two sites 12.6 km apart in Utah at the U.S. Army Dugway Proving Ground.
The HiRes-II detector, located on Camel's Back Ridge, is composed of 
42 spherical mirrors of 3.7 m$^2$ effective area covering nearly 
360$^\circ$ in azimuth.
Half of these, known as ring-one mirrors, cover between $3^\circ$-$17^\circ$
in elevation; the other half (ring-two) cover between $17^\circ$-$31^\circ$ 
in elevation.

Cosmogenic neutrinos, with energies mostly in excess of $10^{18}$ eV, 
are produced via $\pi$ and $\mu$ decays following photopion production from 
high-energy cosmic ray protons incident on the cosmic microwave background 
radiation
\citep{Stecker_PhysRevLett_1968,Margolis.Schramm.Silberberg_ApJ_1978}.
There is evidence to suggest that gamma-ray bursts and active galactic
nuclei jets are possible sources of high-energy cosmic rays and 
neutrinos \citep{Waxman.Bahcall_PhysRevLett_1997,Halzen_ApJ_1997}.
Several theoretical limits on the flux of cosmogenic neutrinos have been 
proposed \citep{Semikoz.Sigl_JCAP_2004,Seckel.Stanev_PhysRevLett_2005}.

Although large uncertainties exist, neutrino cross sections have been 
calculated to vary from $\sim 10^{-32}$ cm$^2$ at $10^{18}$ eV to
$\sim 10^{-31}$ cm$^2$ at $10^{21}$ eV
\citep{Reno_NuclPhysB_2005}.
The opacity of the earth to neutrinos at these high energies therefore 
prohibits the detection of any upward-going event with an elevation angle 
larger than a few degrees.

In the charged-current interaction of a $\nu _e$ in the earth's crust, 
a high-energy electron will be created.
The electromagnetic cascade generated by the electron 
will develop much more slowly due to the onset of the
Landau-Pomeranchuk-Migdal (LPM) effect.
The LPM effect, first described classically by 
\citet{Landau.Pomeranchuk_1953} and later given a quantum-mechanical 
treatment by \citet{Migdal_1956}, predicts that the cross sections for 
bremsstrahlung and pair-production should decrease for a high-energy 
charged particle propagating in a dense medium, 
effectively slowing and elongating the development of the resulting
shower of particles
(a detailed, more modern approach can be found in
\citet{Takahashi_ICRC_2003} and \citet{Baier.Katkov_PhysLettA_2004}).
The energy at which this effect becomes appreciable is inversely 
proportional to the square of the Lorentz factor $\gamma $, and therefore 
the LPM effect should be much more pronounced for the showers generated from 
a $\nu_e$ charged-current interaction than for showers precipitated by 
$\nu_\mu$ or $\nu_\tau$ in the energy range in which HiRes is sensitive.

It is most probable that a neutrino-induced electromagnetic cascade would
be long and nearly-horizontal and observed primarily in the HiRes-II 
ring-one mirrors.
Due to the LPM effect, one expects electron-neutrino-induced showers 
that begin several tens to hundreds of meters deep in the crust to
emerge with enough charged particles to be detected by HiRes-II, thereby
increasing the effective aperture of the detector at high energies.

\section{Search for upward-going neutrino events\label{sec:search}}

The entire HiRes-II data set, which extends from late 1999 to Spring 2006, was
considered when searching for evidence of neutrino-induced upward-going
showers.
Using the standard routines that were developed for analyzing downward-going
cosmic-ray events, we reconstructed the trajectories of each 
upward-going event based on the measured timing and geometry (see 
\citet{Sokolsky_1989} for a description of time- and plane-fitting for 
extensive air showers). 

The data were then filtered in time and position to exclude all calibration 
laser events, which resulted in a loss in the detector aperture of less 
than 1\%.
Additionally and consistent with standard procedure for the analysis of
cosmic-ray data, events were rejected that 
passed within 100 meters of the detector,
had track lengths smaller than 10$^\circ $,
and that had geometrical uncertainties from timing greater than 
36$^\circ $.

\section{The Landau-Pomeranchuk-Migdal Effect\label{sec:lpm}}
At electron energies below the LPM threshold energy ($61.5 \ L_{cm}$ TeV 
\citep{Stanev.Vankov.Streitmatter_PhysRevD_1982}, where $L_{cm}$ is the 
interaction length in cm), the longitudinal profile of an electromagnetic 
shower can be well approximated by the relation

\begin{equation}
  \label{eq:greisen}
  N(t) = \frac{0.31}{{\beta _0}^{1/2}} 
         exp \left[t \left(1 - \frac{3}{2}ln[s] \right)\right] \mathrm{.}
\end{equation}

\noindent
This functional form was first described by \citet{Greisen_1956}, 
with $\beta _0$ as the log of the ratio of the energy of the incident electron 
to its critical energy $E_c$, $t$ as the depth in radiation lengths and 
$s \equiv \frac{3t}{\left[ t + 2 \beta _0 \right]}$.

This relation begins to break down at high energies, greatly underestimating
the distance over which the electromagnetic cascade evolves due to the 
decrease in the cross sections for bremsstrahlung and $e^+e^-$ pair production.
Studies of the electron shower profiles in rock, water and lead above the
LPM threshold energy have been conducted previously 
\citep{Misaki_1990,
Stanev.Vankov.Streitmatter_PhysRevD_1982,
Alvarez-Muniz_ICRC_1999}.
As expected, the results of these analyses show that the shower profiles of 
electron-induced cascades are elongated significantly with respect to the 
Greisen approximation at energies above the LPM threshold, and evolve
differently based on the densities of the media in which the showers propagate.

\section{Calculation of sensitivity to electron-neutrino showers
\label{sec:mcnu}}

To simulate $\nu_e$-induced electromagnetic cascades, we used a four-step 
process.
First, we calculated the average profiles of electron-induced showers using 
the LPM effect.
We then used a Monte Carlo method to simulate the arrival directions and
interaction points of $\nu_e$ around the HiRes detector.
The shower profiles in air were then passed into the HiRes 
detector Monte Carlo to calculate the amount of light seen by the detector.
The HiRes analysis programs were then run on the resulting Monte Carlo 
events to arrive at a $\nu_e$ aperture.

\subsection{Calculating electron-neutrino-induced electromagnetic cascade
profiles\label{sec:simulation}}

In order to treat charged-current $\nu _e N$ interactions in the earth's 
crust, it is necessary to understand the physics of the transition of an 
electromagnetic cascade from a dense medium to a less dense medium (namely, 
from rock to air).
It is therefore important not only to know the number of charged
particles after traversal of a given amount of material in rock, but also the 
energy spectrum of these particles as they leave the ground and enter the 
atmosphere.

We followed the formalism of \citet{Stanev.Vankov.Streitmatter_PhysRevD_1982} 
for calculating the energy-dependence of the probabilities for undergoing pair 
production and bremsstrahlung at LPM energies.
Taking into account any other losses (e.g. Compton scattering and ionization 
energy loss), we calculated two functions:
$N_e ^{rock}(E_0, E, d)$ and $N_e ^{air}(E_0, E, d)$, 
which describe the average number of charged particles with energy $E$ 
resulting from the cascade of an electron, positron or photon with initial 
energy $E_0$ after traversing an amount of material $X$ in rock or air.
The functions $N_e ^{rock}$ and $N_e ^{air}$ were determined for $E_0$ 
at every decade between $10^{12}$ and $10^{21}$~eV using our LPM calculation
and from $E_c$ to $10^{12}$~eV using Equation~\ref{eq:greisen};
LPM calculations of shower profiles from particles with initial energies below 
$10^{12}$~eV were found to be nearly identical to profiles calculated using 
Equation~\ref{eq:greisen}.

\subsection{Simulating neutrino events\label{sec:mc}}

We approximated the earth as a sphere with a radius equal to that at the 
Dugway Proving Ground in Utah.  The density below 58.4~km beneath the surface
(mantle) and the density from from 58.4~km to the surface (crust) were taken 
to be 4.60 and 2.80 g cm$^{-3}$ respectively.
The atmosphere was also simulated up to a height of 50 km above sea level.

Electron-neutrino energies were considered from $\log E_\nu$ of 18 to 21.
The energy-dependence and inelasticity of the charged- and 
neutral-current $\nu N$ interaction cross sections were calculated based on 
the pQCD CTEQ5 model \citep{Lai_EPJC_2000,Gazizov.Kowalski_CompPhysComm_2005}.
From the ratio of the cross sections for charged-current (CC) and 
neutral-current (NC) interactions, 70\% of the events were thrown as 
CC events, while the remaining 30\% were considered NC events.

Neutrino arrival directions were chosen at random such that they only 
penetrated the atmosphere no more than 15$^\circ$ below the horizon.
Events with elevation angles greater than 15$^\circ$ do not contribute 
appreciably to the HiRes-II total $\nu _e$ aperture due to the very small
probability of their transmission through the crust and mantle and subsequent 
interaction near the detector (a 10$^{18}$~eV neutrino at 15$^\circ$ has a 
probability of $\sim 10^{-12}$ of transmission and interaction near the 
detector; this value drops to $\sim 10^{-60}$ at 10$^{21}$~eV).
The variables describing the geometry of the neutrino trajectory were 
determined, such as the distance of closest approach to the detector, the 
vector normal to the shower-detector plane, and the angle of the shower in the 
shower-detector plane.

For these earth-skimming events, the traversal of a critical 
amount of material $X_c$ (measured in g cm$^{-2}$) was found
such that when the shower emerges from the rock into air, it contains
at least $10^7$ charged particles;
showers with a maximum number of charged particles less than $10^7$ will
not trigger the HiRes detector.
This critical pathlength is used to separate the probabilities for 
neutrino transmission and interaction.
The transmission probability $\epsilon _t$, was calculated as the 
probability for a neutrino to penetrate up to $X_c$.
The interaction probability $\epsilon _i$, was calculated from the pathlength
of the neutrino from $X_c$ until escape from the atmosphere.
Since the amount of material traversed in the interaction region is always 
much less than the mean neutrino interaction length, the actual 
point of interaction for each neutrino was then chosen at random for distances 
$X \geq X_c$.
For neutrinos with small elevation angles that do not pass through the Earth,
we considered events that entered the atmosphere above the horizon as well
as those that interacted below the horizon and yielded at least $10^7$ 
particles at the horizon.
For events interacting below the horizon, $X_c$ was taken to be amount of air
penetrated at the horizon.
In the case of events that entered the atmosphere above the horizon, we set 
$\epsilon _t$ to unity and calculated $\epsilon _i$ from the total distance 
traversed in the atmosphere.

\begin{figure}
  \plotone{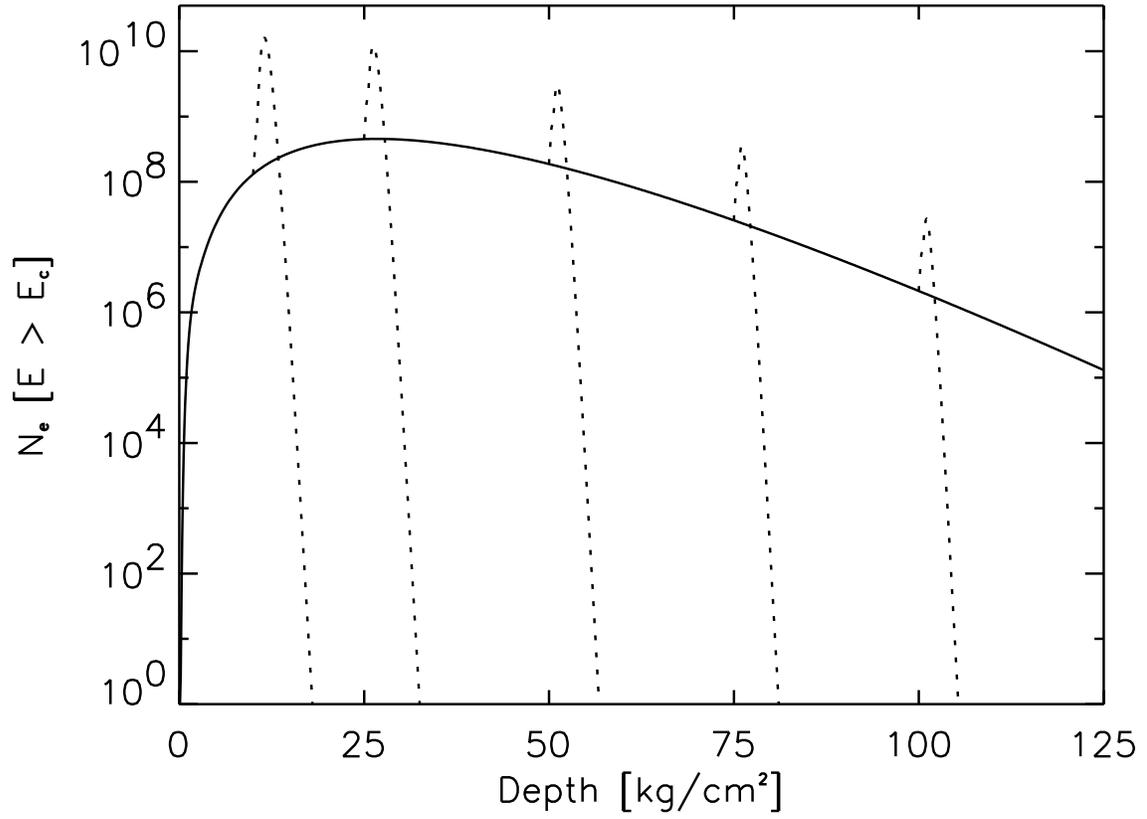}
  \caption{\label{fig:rocktoair}
    An average 10$^{20}$~eV electron shower profile in rock (solid line) with
    average shower profiles for five air showers emerging from the ground at 
    depths of 10000, 25000, 50000, 75000, and 100000 g/cm$^2$ (dashed lines).}
\end{figure}

For all $\nu_e$N interactions the energy transferred to the secondary electron 
or hadron was chosen from the inelasticity distribution ($d\sigma / dy$) 
for the pQCD CTEQ5 model.
For Earth-skimming CC events, the resulting observable profile in 
air was found from a superposition of showers obtained from the energy 
spectrum of electrons, positrons and photons emerging from the rock.
The profiles of CC events that did not pass through the earth were interpolated
from the $N_e ^{air}$ functions described in the previous section.
The profiles for all NC events were calculated using the standard 
Gaisser-Hillas model \citep{Gaisser_1990}.
Each profile was then weighted by a factor 
$w = \epsilon _t \epsilon _i$,
to describe the total probability of transmission and interaction near 
HiRes-II.
Figure \ref{fig:rocktoair} shows the average profiles of five electron-induced 
air showers emerging from the ground at different depths along an average 
$10^{20}$~eV electron-induced shower in rock.

\subsection{Simulating detection by HiRes}

Having generated shower profiles using the LPM effect, the shower 
profiles were then passed through a HiRes Monte Carlo program
which models the response of the detector to cosmic-ray-induced showers.
This program determines the amount of fluorescence and \v{C}erenkov 
photons produced for a given number of charged particles, and scatters 
and attenuates the light appropriately when given the known variables 
describing the geometry of the shower with respect to the detector.
The program then models the HiRes-II trigger conditions to decide if the
simulated shower is read out by the detector \citep{Abbasi_PRL_2004}.

\subsection{Analysis and filtering of simulated events}

Simulated showers which triggered the detector were analyzed using the 
same routines used in the analysis of the real data used in our
search for neutrinos in the upward-going HiRes-II data.
The variables describing the geometry of the shower were fit and compared
to the known variables.
An event was considered accepted when it passed the same cuts described in 
Section 2.

\section{Calculating an electron-neutrino flux upper limit
\label{sec:upperlimit}}

\begin{figure}
  \plotone{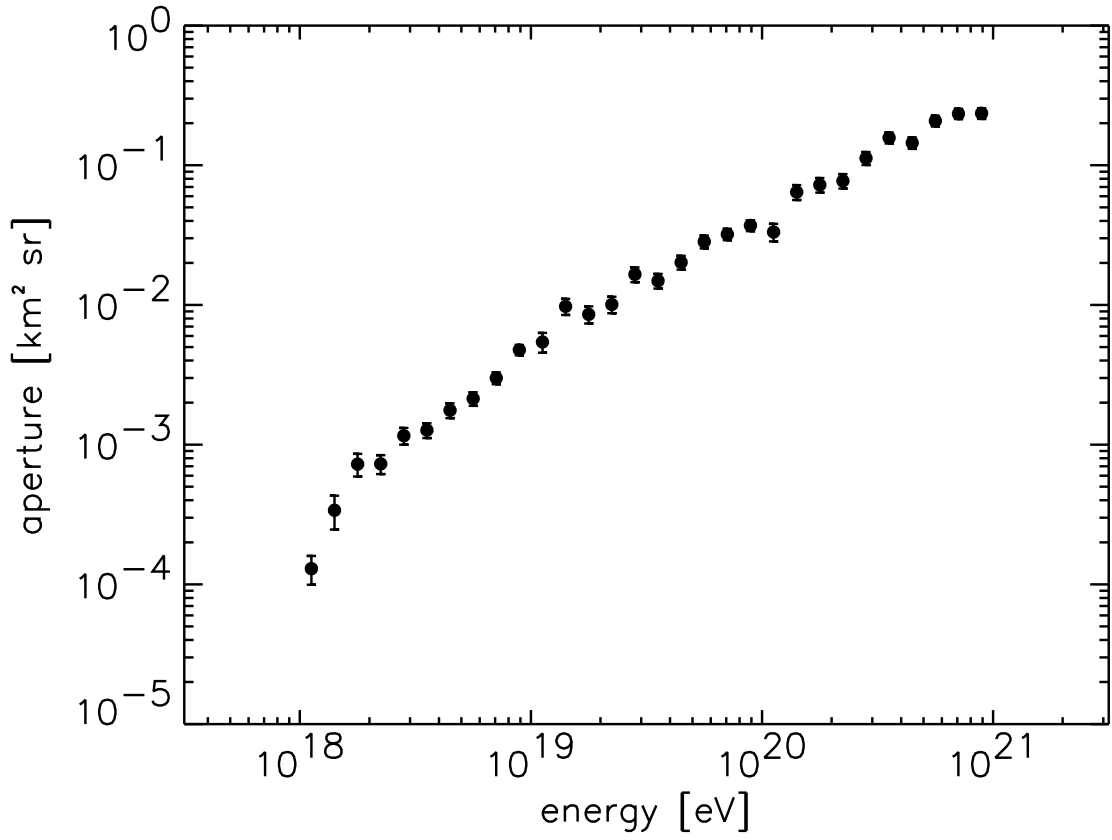}
  \caption{\label{fig:aperture}
    The calculated HiRes-II electron-neutrino aperture.}
\end{figure}

For the purposes of arriving at a predicted HiRes-II $\nu _e$ aperture, the 
simulated events were collected in 30 0.1-decade energy bins from 
$10^{18}$ to $10^{21}$~eV.
The aperture for a given energy bin was found

\begin{equation}
  \label{eq:aperture}
    \left( A \Omega \right)_E =
    \left[ 2 \pi \int _{0 ^{\circ}} ^{30 ^{\circ}}
    sin (\theta) \ d\theta \right]^2
    R^2 \left( \frac{\sum _i ^{N_a} w _i}{N_T} \right)_E 
\end{equation}

\noindent
where $R$ is the radius of the earth extended 50 km to the edge of the 
atmosphere.
The geometrical component of the aperture is derived from the area 
and solid angle of a $30^\circ$ cap on a sphere of radius $R$.
This is then adjusted by the weighting factor $w$ (discussed in Section 4.2) 
for each of the $N_A$ events that trigger the detector out of a total 
$N_T$ events thrown in the given energy bin.
The HiRes-II $\nu _e$ aperture is shown in Figure~\ref{fig:aperture}.

Consistent with our study of $\nu_\tau$
\citep{Martens_2007},
we calculate a flux limit in three energy bins:
$\Delta E~=~10^{18} - 10^{19}$, 
    $10^{19} - 10^{20}$, and 
    $10^{20} - 10^{21}$~eV,
over the total HiRes lifetime of 3638 hours.
We observe no neutrino events over the entire energy range.
We calculate the flux limit 
($E^2 \frac{dN}{dE}$)
at the 90\% confidence level to be 
$4.06 \times 10^3$, 
$3.55 \times 10^3$ and
$4.86 \times 10^3$ eV cm$^{-2}$ sr$^{-1}$ s$^{-1}$ at
$10^{18.5}$, 
$10^{19.5}$ and 
$10^{20.5}$~eV, respectively.
Combined with our $\nu_\tau$ results and provided equal mixing of all
neutrino flavors, this reduces the limit to
$3.81 \times 10^2$,
$9.73 \times 10^3$ and
$4.71 \times 10^3$ eV cm$^{-2}$ sr$^{-1}$ s$^{-1}$. 

\section{Discussion\label{sec:discussion}}

As is the case with all high-energy neutrino calculations, the largest
uncertainty lies in the extrapolation of $\nu N$ cross sections.
Different cross section models can cause the limits to vary somewhat.
The incorporation of cross sections from previous and more recent versions 
of the CTEQ model can change the limits by as much as 10 to 40\% at the 
lowest and highest energies, respectively.

Recent work imposing the Froissart bound on structure functions for 
extrapolating $\nu N$ cross sections show a decrease in cross sections at
$10^{21}$ eV by about a factor of 8 over the CTEQ5 
parameterization \citep{Block_2007}.
These cross sections increase our $\nu _e$ limit by 40\% at
the lowest energy bin and increase the value of our highest energy bin 
by a factor of ~3.

In addition to uncertainties in $\nu N$ cross sections, our limits are also
sensitive to the energy transferred to the secondary electron or hadron.
From parameterizations of the mean inelasticity in $\nu N$ interactions 
\citep{Quigg.Reno.Walker_PhysRevLett_1986}, if we allow the transfer of 
exactly 80\% of the neutrino energy to the electron (and 20\% to the hadron), 
our $\nu _e$ limits will increase between about 15\% at $10^{18.5}$~eV, remain
unchanged in the middle energy bin and decrease by about 5\% at $10^{21.5}$~eV.

\section{Conclusion\label{sec:conclusion}}

We have found no evidence of upward-going neutrino-induced cosmic-ray showers
in the HiRes-II data.
We have presented a technique for modeling the full HiRes-II detector response
to ultrahigh energy neutrino-induced LPM cascades in rock and air.
With no neutrino events seen in the HiRes-II data, and provided equal mixing 
of all neutrino flavors, we have found an upper limit on the flux of 
ultrahigh energy neutrinos at a 90\% confidence limit.

\begin{figure}
  \plotone{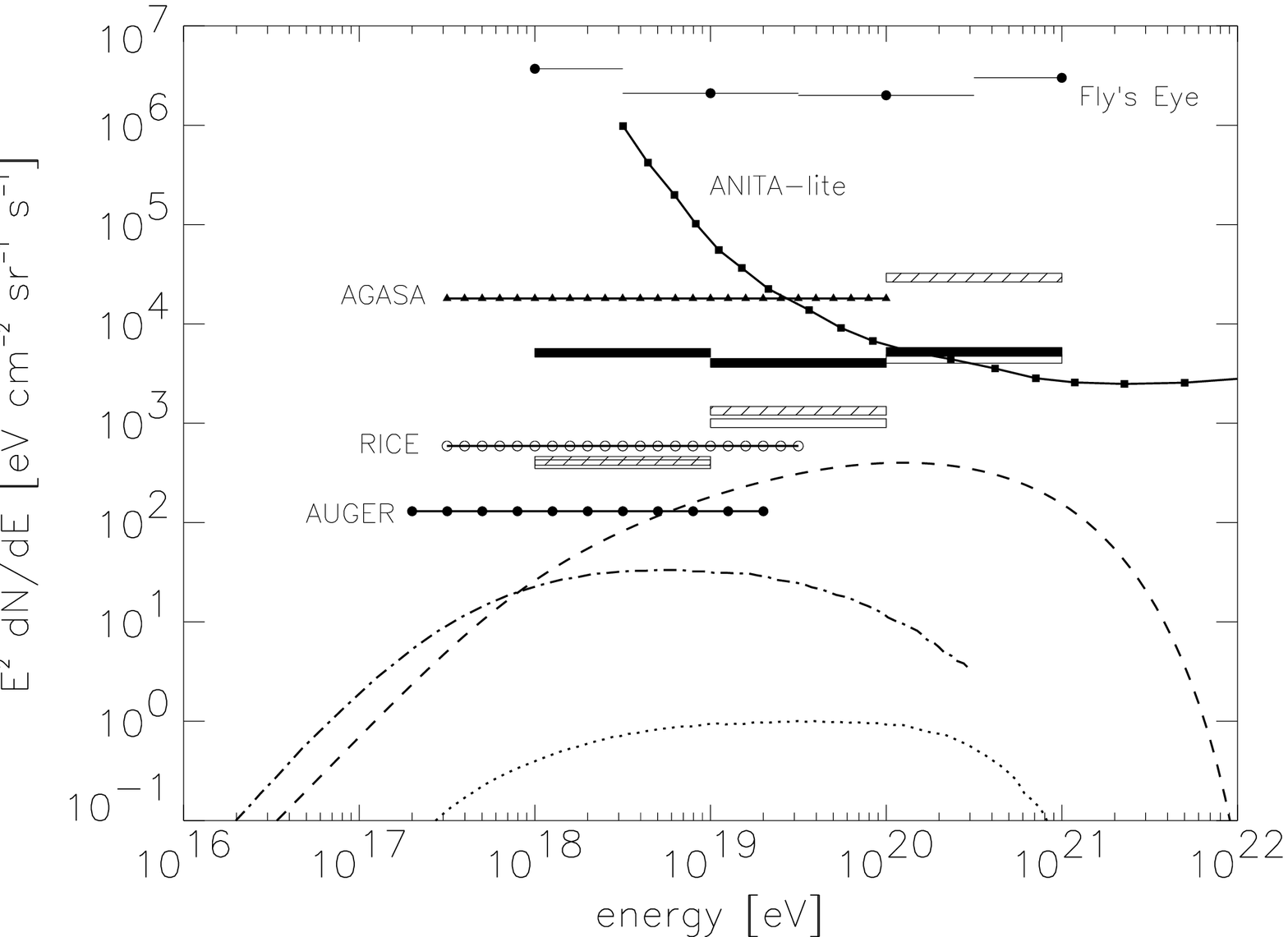}
  \caption{\label{fig:flux_limit}
    The HiRes-II neutrino flux limit.
    \emph{black boxes}: $\nu_e$ limit (this work).
    \emph{cross-hatched boxes}: $\nu_\tau$ limit \citep{Martens_2007}.
    \emph{open boxes}: $\nu_e$ and $\nu_\tau$ combined flux limit.
    \emph{Dotted line}: cosmogenic per flavor neutrino flux limit from fits
    to HiRes cosmic-ray data \citep{Brusova_2007}.
    \emph{Dashed line}: cosmogenic per flavor neutrino flux limit derived
    from fits to existing cosmic- and gamma-ray data 
    \citep{Semikoz.Sigl_JCAP_2004}.
    \emph{Dot-dashed line}: cosmogenic per flavor neutrino flux from fits
    to HiRes and AGASA cosmic-ray data \citep{Seckel.Stanev_PhysRevLett_2005}.
    Also shown are calculated neutrino flux limits from the
    Fly's Eye \citep{Baltrusaitis_ApJ_1984,Baltrusaitis_PhysRevD_1985},
    ANITA-lite \citep{Gorham_PhysRevLett_2004}, 
    RICE \citep{Kravchenko_PhysRevD_2006},
    AGASA \citep{Yoshida_ICRC_2001} and
    Auger \citep{Abraham_PRL_2008} experiments.}
\end{figure}

Figure \ref{fig:flux_limit} shows the upper limit on the neutrino flux 
from the analysis of the HiRes $\nu _e$ and $\nu _{\tau}$ flux limits as
compared to three theoretical curves and to calculated flux limits from other
experiments.
The $\nu_e$ flux limits reported here have improved upon those 
for the Fly's Eye by about two and a half orders of magnitude.
Combined with the results of the $\nu_\tau$ analysis, this limit lies just
above the theoretical neutrino flux of \citet{Semikoz.Sigl_JCAP_2004}, and 
about an order of magnitude above that of
\citet{Seckel.Stanev_PhysRevLett_2005}.
Our combined neutrino flux limit is about two and a half orders of magnitude 
above the cosmogenic neutrino flux predictions of \citet{Brusova_2007}, which 
has been derived from a proton injection model with cosmologically evolving 
sources and injection spectra that fit the HiRes cosmic-ray spectrum.

\acknowledgements
We would like to thank Steve Barwick for useful discussions and 
recommendations while writing this paper.
This work was supported by US NSF grants PHY-9100221, PHY-9321949,
PHY-9322298, PHY-9904048, PHY-9974537, PHY-0073057, PHY-0098826,
PHY-0140688, PHY-0245428, PHY-0305516, PHY-0307098, PHY-0649681, and
PHY-0703893, and by the DOE grant FG03-92ER40732.  We gratefully
acknowledge the contributions from the technical staffs of our home
institutions. The cooperation of Colonels E.~Fischer, G.~Harter and
G.~Olsen, the US Army, and the Dugway Proving Ground staff is greatly
appreciated.

\end{document}